\documentclass[traditabstract]{aa} 
\usepackage{graphicx}
\usepackage{txfonts}
\usepackage{wasysym}
\usepackage{natbib}
\bibpunct{(}{)}{;}{a}{}{,} 
\usepackage{ulem}
\begin{document}
   \title{TASTE: The Asiago Search for Transit \\ timing variations of Exoplanets}
     \subtitle{I. Overview and improved parameters for HAT-P-3b and HAT-P-14b} 
   \author{V. Nascimbeni\inst{1,2}\thanks{Visiting PhD Student at STScI under the DDRF D0001.82432
           program.}
          \and G. Piotto\inst{1}
          \and L.\ R. Bedin\inst{2} \and M. Damasso\inst{1,3}
          }

   \institute{Dipartimento di Astronomia, Universit\`a degli Studi di Padova,
              Vicolo dell'Osservatorio 3, 35122 Padova, Italy\\
              \email{valerio.nascimbeni@unipd.it, 
                     giampaolo.piotto@unipd.it, 
                     mario.damasso@studenti.unipd.it}
         \and
             Space Telescope Science Institute,
             3700 San Martin Drive, Baltimore, MD 21218\\
             \email{bedin@stsci.edu}
         \and
             Astronomical Observatory of the Autonomous Region of the Aosta Valley, 
             Loc.\ Lignan 39, 11020 Nus (AO), Italy
             }

   \date{Accepted 2010 November 25}

\abstract{ A promising method for detecting earth-sized exoplanets is the
  timing analysis of a known transit. The technique allows  a search
  for variations in either the transit duration or the center induced by the
  perturbation of a third body, e.g.  a second planet or an
  exomoon. By applying this method, the TASTE (The Asiago Search for Transit Timing
  variations of Exoplanets) project will collect high-precision,
  short-cadence light curves for a selected sample of transits by
  using imaging differential photometry at the Asiago 1.82m
  telescope. The first light curves show that our project can 
  achieve a competitive timing accuracy, as well as a  significant
  improvement of the orbital parameters. We derived refined
  ephemerides for HAT-P-3b and HAT-P-14b with a timing accuracy
  of 11 and 25 s, respectively.}

   \maketitle

%
\section{Introduction}
\label{intro}
%

Among the  techniques developed  to discover  exoplanets, the
photometric transit  method is one of the most promising having already
discovered more than 100  planets.  Transits also give us the unique
opportunity of a nearly complete characterization of an extrasolar
planetary system, by measuring orbital and physical parameters that
would otherwise be inaccessible: the planetary radius $R_{\rm p}$,
the orbital inclination $i$, the ``real'' planetary mass $M_{\rm p}$
(which cannot be provided by the radial velocities alone), and hence
the average density ($\rho_{\rm p}$). 

The great opportunities provided by transits however exist with some
disadvantages.  In
order to obtain a light curve  of a typical target  with a significant
signal-to-noise ratio (S/N), an  extremely high photometric precision
needs to be reached and sustained for  several hours.  Many systematic
trends can arise from various sources.  Known as \emph{correlated} or
\emph{red noise}, this effect plagues to a variable extent 
the ground-based photometry in particular \citep{pont2006} and specific techniques
have been developed to deal with it (e.g., Tamuz et al. 2005).
\nocite{tamuz2005}   The simultaneous requirement of both precision
and stability makes a sub-millimagnitude light curve a real challenge
for ground-based instruments. 

In spite of these difficulties, even exoearths  might be within the
reach of ground-based facilities.  Several indirect methods  for
detecting them have been proposed.  One of these is \emph{Transit
  Time Variation} (TTV) analysis.  In principle, a single planet on a
Keplerian orbit should be seen to transit at very regular time
intervals. However, if there were  a second planet (not necessarily
transiting) in the same system, it would gravitationally perturbate
the orbit of the  transiting  one, breaking its strict periodicity and
varying the measured mid-transit time relative to the expected one
\citep{holman2005}.  The amplitude of the TTV is  dependent upon the
mass of the perturber (more massive planets leading to a larger
effect) and is strongly enhanced if the hidden planet is locked in a
low-order orbital resonance with the primary, such as 1:2 or 2:3
\citep{agol2005}.  In that case, even a terrestrial planet could be
responsible for a TTV on the order of tens or hundreds of seconds.  It
should be noted that such resonant orbits have already been detected
with the radial velocity (RV) method in some planetary systems
\citep{tinney2006}.

Many ground-based projects about TTV analysis are currently ongoing.
The most notable are RISE\footnote{\sf
  http://telescope.livjm.ac.uk/Info/TelInst/Inst/RISE/}
\citep{steele2008}, which  published null  TTV detections of TrES-3b
and HAT-P-3b \citep{gibson2009,gibson2010},  and The Transit Light
Curve Project \citep{holman2006}.  Most of these works use an
instrumental setup that is specifically designed to get
high-precision light curves with a  short cadence,  sometimes of a few
seconds.  A very high sampling rate is required to 
accurately determine the instant at which the mid-transit occurs.  Systematic trends have to be
carefully corrected, because they can  perturb the light curve from
its ideal symmetric shape, biasing the estimated timing and, possibly,
mimicking a TTV  \citep{gibson2009}.   Despite the efforts of
these and other authors, achieving a timing accuracy $\lesssim$10 s is
still an ambitious target, typical values ranging from 15 to 30 s for
the most accurate published works. 

Interestingly, even though five years  have elapsed since the first TTV study
\citep{steffen2005}, no clear evidence has so far been found for a
genuine perturbation by a previously unknown planet, although we note
that Holman et al. (2010) announced the Kepler discovery of a pair
of giant planets orbiting the same star in a 2:1 resonance, and
showing an impressive TTV signature (120--240 s). \nocite{holman2010}
Some ambiguous results have instead been reported for OGLE-TR-111
\citep{pont2004,diaz2008} and a (yet unconfirmed) claim for WASP-3b
\citep{pollacco2008} was published by \citet{mac2010}.  Perhaps  small
planets in orbital resonances are less common than previously expected
by migration models such as that of \citet{cresswell2006}, which would be by
itself a significant result.  We note that the timing
precision reached by most of the earlier studies  was around 50--100
s, and that the effects of red noise were then more poorly understood than they are
today.

Compared to TTV, \emph{Transit Duration Variation} (TDV) analysis is a
more recently developed method, whose theory is more subtle.  A periodic  change
in the transit duration may arise from the presence of a satellite, or
\emph{exomoon}, which makes the planet oscillate around the
planet-satellite barycenter along its path \citep{kipping2009a}; in
addition, a periodic TTV signal is expected for the same reason.  Here
TTV and TDV are strictly complementary: a $\pi /2$ phase difference
between the two signals is the expected distinctive signature of an
exomoon, and the ratio of their amplitudes allows the satellite mass and its distance
from the planet to be determined separately \citep{kipping2009b}.  For
the most part, timing surveys are focused only on the TTV. A tricky
problem  in determining the  duration is that it is very dependent
upon the limb darkening (LD) parameters, which are  poorly
constrained.  Moreover, the LD is wavelength-dependent, making the
comparison of TDV data extracted with different instrumental setups
prone to systematics.

Our primary aim is to collect a database of high-precision light
curves that will be suitable for a simultaneous TTV/TDV analysis,
optimizing every task from the observation/calibration setup to the
data extraction with fully home-made  software tools, optimized for
this specific program. Our survey is based on data collected  at the
Asiago Observatory\footnote{\sf http://www.pd.astro.it/asiago/}.
Some feasibility tests were also performed at the Osservatorio
Astronomico della regione autonoma Valle d'Aosta\footnote{\sf
  http://www.oavda.it/} (OAVdA) observatory, which hosts a 0.81m
telescope as  its main  instrument. Once the necessary fine tuning
are completed, it will be useful as an auxiliary facility for our
program, improving the efficiency of our survey.

In this first paper of this series, we present preliminary results,
describing in detail: the strategy, facility, 
observations, data reduction, and analysis.  We  demonstrate that
---even at this early stage--- our photometry and time resolution are
competitive compared with those derived by instruments of the
same class.  Two targets of our survey  are shown as examples, and  we
present improved  orbital and physical parameters of the recently
discovered HAT-P-14b \citep{torres2010} and of HAT-P-3
\citep{torres2007},  based on our new data.

\section{Target selection}
\label{sample}

\begin{table*}
\caption{
  The selected sample of targets for the TASTE project, and their parameters.}
\label{effects}
\centering
\scalebox{0.88}{
\begin{tabular}{llllllllllllll}
\hline\hline
name&         $M_{\rm p}/M_J$ & $R_{\rm p}/R_J$  & $P$ (d)& dur (min) & $i$ ($^\circ$)&  RA & DEC & $V$ & depth & TTV$_{2:1}$ (s) &    TTV$_{\rm m}$ (s)&       V-TDV$_{\rm m}$ (s) &    T-TDV$_{\rm m}$ (s) \\
\hline
 WASP-11        &0.46	&1.05	&3.72	&159	&88.50	&030929	&+304025 	&11.9	&0.018 &485	&4.13	&8.17	&0.06\\
 HAT-P-9        &0.78	&1.40	&3.92	&206	&86.50	&072040	&+370826	&12.3	&0.012 &302	&2.67	&6.43	&0.18\\
 XO-5  	        &1.08	&1.09	&4.19	&193	&86.80	&074652	&+390541	&12.1	&0.011 &234	&2.59	&5.51	&0.17\\
 XO-2  	        &0.57	&0.97	&2.62	&162	&88.58	&074807	&+501333	&11.2	&0.011 &276	&2.38	&6.81	&0.02\\
 HAT-P-13       &0.85	&1.28	&2.92	&194	&83.40	&083932	&+472107	&10.6	&0.007 	&206	&1.89	&5.80	&0.35\\
 HAT-P-3        &0.60	&0.89	&2.90	&123	&87.24	&134423	&+480143	&11.9	&0.012 	&291	&2.60	&5.10	&0.12\\
 HAT-P-12       &0.21	&0.96	&3.21	&140	&90.00	&135734	&+432937	&12.8	&0.020 	&906	&6.22	&12.54	&0.00\\
 HAT-P-14       &2.23	&1.15	&4.63   &131    &83.50	&172028 &+381432        &9.98   &0.006  &125	&1.52	&1.99	&1.23\\
 HAT-P-5        &1.06	&1.26	&2.79	&175	&86.75	&181737	&+363718	&12.0	&0.012	&158	&1.59	&4.61	&0.08\\
 WASP-3         &1.76	&1.31	&1.85	&137	&85.06	&183432	&+353942	&10.6	&0.011 	&63	&0.74	&2.52	&0.05\\
 HAT-P-1        &0.52	&1.23	&4.47	&160	&86.28	&225747	&+384030	&10.4	&0.013  &511	&4.10	&6.77	&0.41\\
 HAT-P-6        &1.06	&1.33	&3.85	&203	&85.51	&233906	&+422758	&10.5	&0.009	&219	&2.13	&5.16	&0.22\\
\hline
\end{tabular}}
\tablefoot{The first ten columns give the name of the
target, the planet mass $M_{\rm p} $ and radius $R_{\rm p}$ (both in
jovian units), the orbital period $P$ (in days), the transit duration
(in minutes), the orbital inclination $i$ (in degrees), the 2000.0 RA
and DEC, the $V$ magnitude of the host star, the transit depth (in
flux). In the last four columns, we reported the expected TTV \& TDV
effects in seconds caused by a 1 $M_\oplus$ exoplanet in a 2:1 resonant
external orbit (TTV$_{2:1}$) and to a 1 $M_\oplus$ exomoon at one
third of the Hill radius (TTV$_{\rm m}$ and the V- and TIP- components
of the TDV$_{\rm m}$).}
\end{table*}

For each studied exoplanet, a TTV/TDV analysis requires the
construction of a homogeneous database of  photometric high-precision
light curves (RMS: $\sim$1 mmag), all calibrated in time, at a
high-accuracy level ($\lesssim$1 s). Thus, it is necessary to perform
a careful selection of the targets, based on both observational and
physical criteria.  Among the 81 transiting exoplanets known at the
time of the submission  (\texttt{exoplanet.eu} database), we assigned
a rating to each candidate by taking into account the following
observational constraints:
\begin{itemize}
\item declination $\delta > +0^\circ$;
\item magnitude of the host star $9<V<13$;
\item orbital period $P\lesssim 5$ days;
\item total duration $d\lesssim 200$ minutes;
\item transit depth $\gtrsim 5$ mmag;
\item the presence of suitable reference star(s)  within the field of
  view of our detectors. 
\end{itemize}
As demonstrated by our tests on the Asiago 1.82m telescope, this first
set of requirements gives us the chance to collect each season a
reasonable number of full transits with a high sampling rate and a high
S/N. 

Targets lying within the CoRoT or Kepler fields were excluded; targets
discovered more than four years ago, or for which accurate TTV/TDV studies have
been already published, were assigned a lower ranking.  The last two
criteria take  into  account the probability of a large TTV/TDV
effect (both periodic or secular) being detected by space missions or
previous ground-base studies. In addition, we evaluated other
factors specific to the detectability of the TTV/TDV signal:
\begin{itemize}
\item the computed TTV$_{2:1}$ effect induced by a 1 $M_\oplus$
  exoplanet  in a 2:1 resonant external orbit \citep{agol2005};
\item the computed TTV$_{\rm m}$ effect induced by a 1 $M_\oplus$
  exomoon  at one third of the Hill radius \citep{kipping2009a}, which
  can be conservatively assumed to be the allowed region for a stable
  satellite;
\item the computed TDV$_{\rm m}$ effect induced by the same exomoon,
  for both the V (velocity change) and the TIP (impact parameter)
  components of the perturbation discussed by \citet{kipping2009b}.
\end{itemize}
Targets with TTV$_{2:1}<30$ s were excluded. Among the remainder, a
higher priority was set for targets with a total TDV$_{\rm m}>5$ s.
We briefly discuss in Section \ref{ttvan} how a TTV/TDV signal will be
detected.

Finally, a sample of twelve exoplanets was selected.  The list sorted
in order of RA, along with a summary of their parameters can be seen
in Table \ref{effects}.    
Few targets were added  to the list, as backup choices or  because of
their interesting properties, though they  do not fully comply with
the selection criteria.  For instance, HAT-P-14b is  quite a massive
planet, but its highly ``grazing'' configuration makes it  very
sensitive to changes in the transit duration. On the other hand, two
targets were included because a TTV is expected by previous studies:
the above-mentioned WASP-3b and HAT-P-13b, for which a TTV is
predicted by the perturbation of a known -- but non-transiting --
outer planet (Bakos et al. 2009). \nocite{bakos2009}

\section{Instrumental setup and observational strategy}
\label{setup}
%

The facility used for the present work is the Asiago Astrophysical
Observatory, located at Mount Ekar (elevation: 1366 m) in northern
Italy.  Its 1.82m Cassegrain is the largest optical telescope within
Italian territory.

The instrument for imaging at the 1.82m is the Asiago Faint Object
Spectrograph and Camera (AFOSC),  a focal-reducer type camera.  Its
9$'$$\times$9$'$ field of view is large enough to allow a careful
selection of the stars to be used as reference sources in differential
photometry.  When used as an imager, AFOSC is equipped with a set of
standard Bessel $UBVR$ filters, and a Gunn $i$ filter.

The CCD detector was recently upgraded, by installing a 2k$\times$2k
pixels thinned, back-illuminated E2V 42-20, which provides a
$\sim$90\% quantum efficiency in the $R$ band spectral region. Its
pixel size implies a $0.26''/$pixel scale, which severely oversamples
the stellar profiles considering the typical seeing of the
site. Despite the thermoelectrical cooling, usually set at $-$60
$^\circ$C, the measured dark current is quite low, peaking at 0.2
e$^-$/s/pix in the hottest regions. The cosmetics are very good, with
only a bad column starting from a hot pixel group.  The 500 MHz
read-out mode provides a 2.7 e$^-$/ADU  gain and a $\sim$7.6 e$^-$
read-out noise (RON), which may seem quite high compared to the
current standards but does not  represent an obstacle for our
project because it is in a shot-noise (and scintillation-) limited
regime   (as we later illustrate). 

The camera controller software has been  customized  to meet our needs
for a very high sampling rate. Binning factors of up to 4$\times$4 are
available, which help us to lower the readout time while at the
same time decreasing the readout noise. For the same reason, sub-windowing is
used when only a part of the field is necessary to image a suitable
set of reference stars. Significant effort has been devoted to shortening as
much as possible the dead technical time between consecutive
frames. The total overhead is now 3.4 s when reading a full frame,
and 2.2 s for one quarter of a 1024$\times$1024 frame.  This performance
allows us to achieve a sampling of about 7 s and 5 s, respectively, while keeping a
duty-cycle $>$50\%.  

We ran tests of the accuracy of the mechanical shutter.  For exposures
as short as 1s, the effective integration time is repeatable to within
0.05 s and the  travel time is regular to about two parts in $10^{4}$,
allowing sub-millimagnitude photometry.  However, we do not wish
to reach such short exposures, since the duty-cycle would
get very low ($<$20\%) and a non-negligible fraction of the overall
S/N of our light curve would be lost, nullifying the advantages of a
fast sampling. When a target is too bright to be imaged with
$\ge 2$ s exposures, we choose instead to apply a slight defocus  to avoid
saturation.  The defocusing technique also minimizes systematic errors
caused by a non-perfect flat field correction
\citep{southworth2009b,southworth2009a}.  The 1.82m autoguide is not
accurate enough to keep the stars on the same physical pixel during
the whole series, but our data analysis will take into account the
motion of the centroids as a  possible source of systematic errors.

Only a good  \emph{pre-reduction strategy}  would guarantee the
extremely precise  photometry required by the goals of the project,
and much of the efforts are devoted to this task.  Even very
small instrumental drifts (over a timescale of just a few hours)
could easily reach amplitudes of  a few mmag.  Bias, dark, and flat-field 
frames are taken just before and after the photometric series.
In this way, at the data-reduction stage   it becomes possible to
linearly interpolate between the pre- and post- master-bias, -flats, and
-darks to ``customize'' the calibration frames for each individual
scientific frame of the time series.  The master-bias, master-flat, and
master-dark frames collected on each night are also stored for a
long-term calibration analysis, in order to highlight any change or
drift that may be correlated with seasonal trends (such as temperature)
or the detector decay (cosmetics). 

\begin{table*}
\caption{Summary of the observed transits at the 1.82m Asiago telescope.} 
\label{observ}
\centering
\begin{tabular}{lllllllll}
\hline\hline
target   & date        & UT obs. time  & airmass     & filter & exptime (s)  & sampling (s)  & frames & notes   \\ \hline
XO-3     & 2010 Feb 13 & 19:03$-$22:08 & 1.05$-$1.29 & $R$    & 2, 1.5, 1 & 5$-$6   & 2238   & partial, clouds \\
WASP-12  & 2010 Feb 14 & 20:36$-$23.59 & 1.04$-$1.28 & $R$    & 2, 1.5    & 5$-$6   & 2310   & partial  \\
HAT-P-14 & 2010 Mar 12 & 01:25$-$04:53 & 1.34$-$1.01 & $R$    & 2         & 5.4        & 2247   & complete \\
HAT-P-3  & 2010 Apr 07 & 23:23$-$03:31 & 1.00$-$1.24 & $R$    & 2         & 5.0        & 2882   & complete \\ \hline
\end{tabular}
\end{table*}

Even when taking the greatest care, smooth gradients on the order of
$\lesssim$5\% are evident when comparing dome flats with a more
accurate sky flat.  Therefore, sky flats, taken at
both the sunset and dawn twilights and then interpolated, provide a 
more effective means of flat-fielding our data. Dome
flats are taken in any case, not only as a backup means of performing the flat-field
calibration, but also to check for short-period variations in
sensitivity. Even when dealing with the highest quality flats, accuracy at low
spatial frequencies is challenged by ``sky concentration'', 
typically affecting focal reducers \citep{andersen1995} where scattered light
is directed preferentially at the center of the field.  As we 
later show, a very smooth gradient is not a problem as long as the
tracking drifts are carefully monitored and constrained to within a few
pixels.

%
\section{Observations}
\label{observs}
%

\begin{figure*}
\centering
\framebox{\includegraphics[height=8cm]{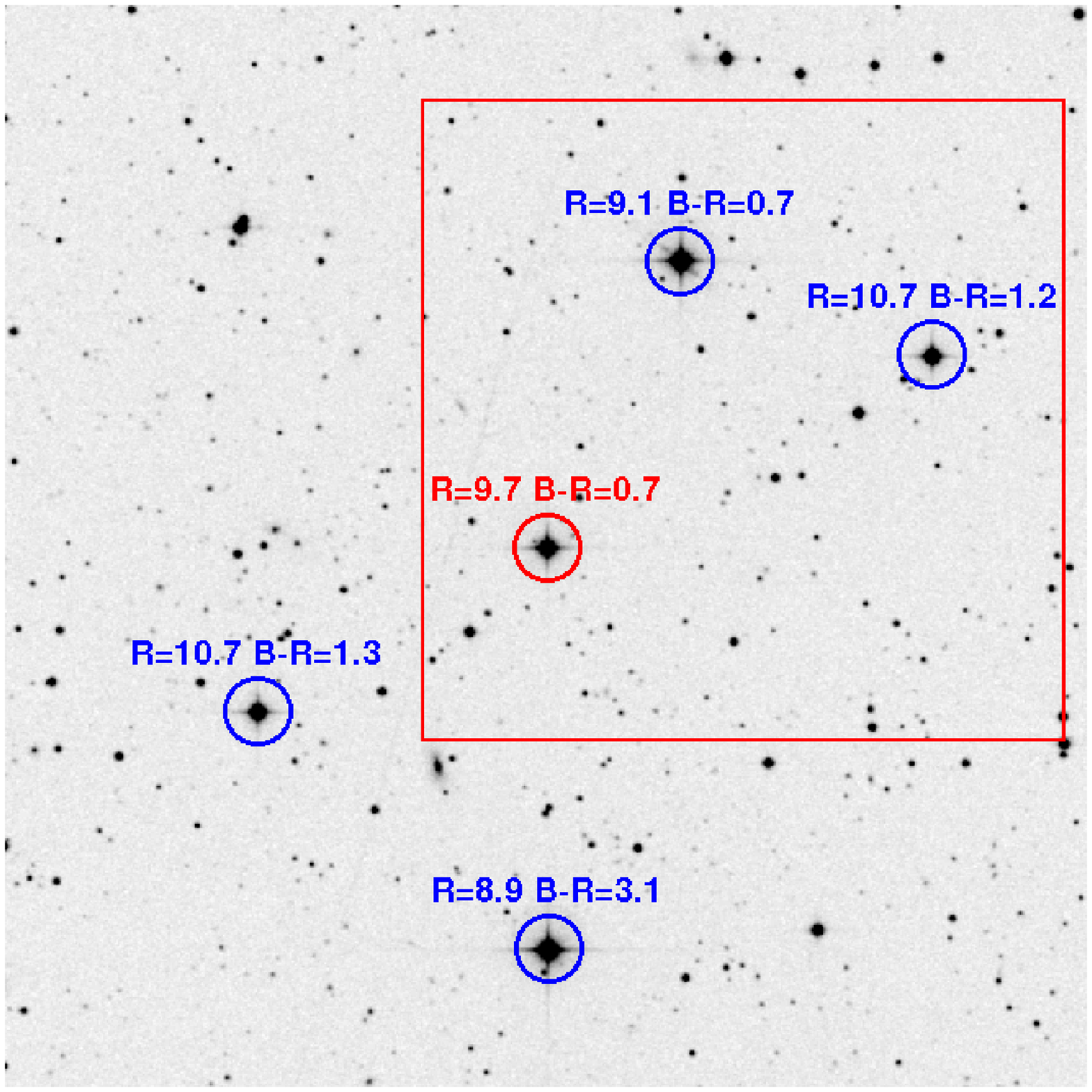}}\hspace{0.5cm}
\framebox{\includegraphics[height=8cm,clip=true,trim = 7 10 7 10]{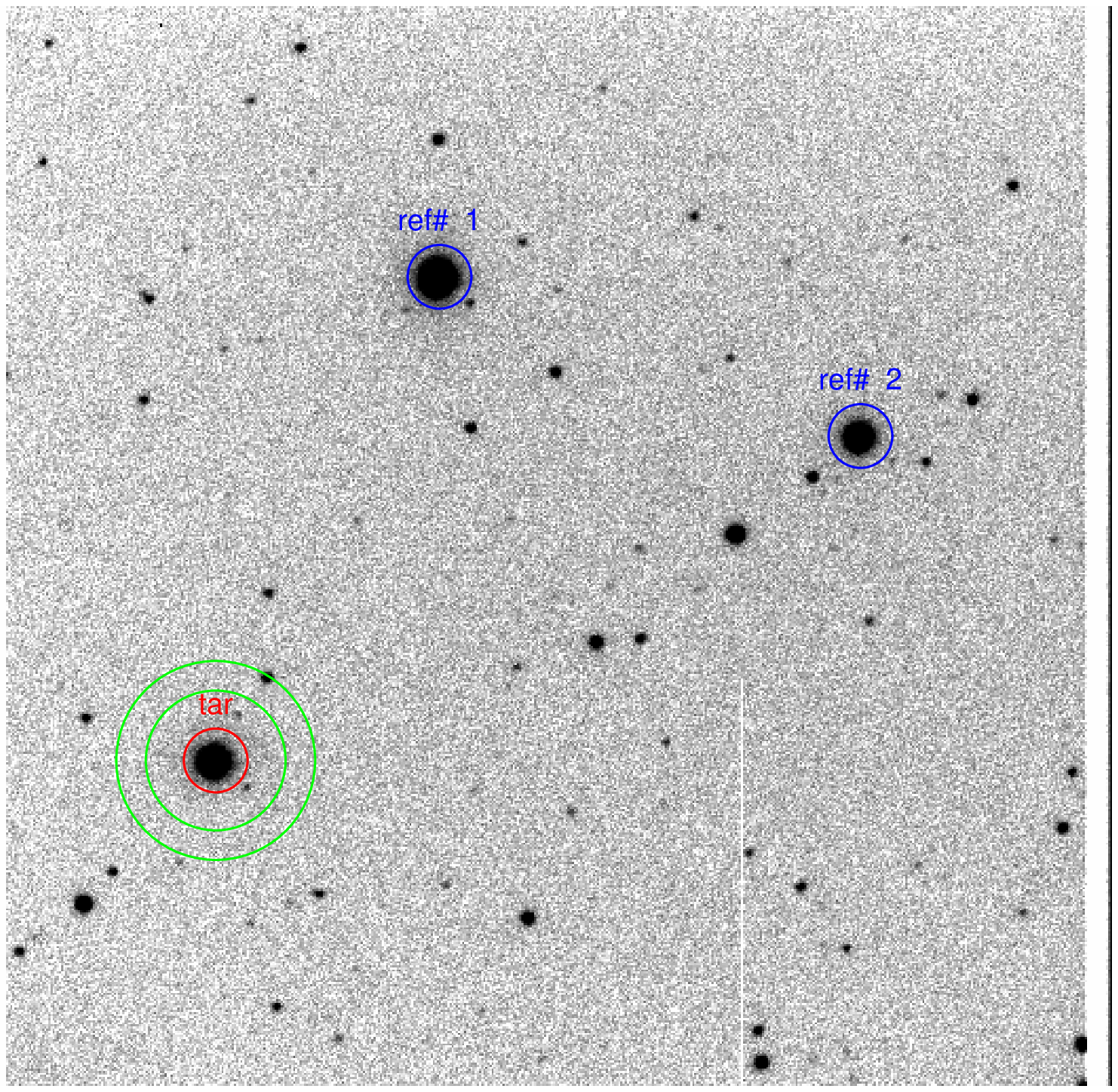}}\\
\caption{
(\textit{Left}): 
A 15$'$$\times$15$'$ image
from PSS2-red centered on HAT-P-14 (red circle) 
North is up, East is to the left. For reference, 
the blue circles highlight the brightest stars in the field (corresponding
numbers give their $R$ magnitude and $B-R$ color).
The AFOSC full-frame FOV (red square) is also shown. 
(\textit{Right}): 
Example of the typical image quality obtained with AFOSC 
on a 2 s single frame during the 2010 Mar 12 run. The
two main reference stars are marked with blue circles. 
The two green circles indicate the annulus within which 
the local sky was estimated. 
}
\label{charts}
\end{figure*}

After a tuning phase that lasted some months mostly involving
hardware/software tests and closed-dome calibrations, the first
scientific observations on real transits  were performed in February
2010, namely for XO-3b \citep{krull2008} on 2010 February 13, and WASP-12b
\citep{hebb2009} on 2010 February 14\footnote{Dates for the observing
  nights are reported as  ``evening dates''.}. In both cases, clouds
and high humidity prevented a full transit from being
monitored. However, a data analysis of that light curves revealed some
problems (such as dome turbulence, hardware/software bugs) for which
later countermeasures were taken.  All the  observations reported here
were carried out with the AFOSC $R$ filter.  The log of the
observations shown in Table \ref{observ}.

The first complete light curve was measured for HAT-P-14b on the night
of 2010 March 12. HAT-P-14b is a ``hot Jupiter'' 
discovered by \citet{torres2010},  orbiting a slightly evolved F dwarf
($V = 9.98$).  This planet is more massive and eccentric than the
average of its class.   In addition, its grazing inclination generates
a V-shaped light curve, which differs from the usual U-shaped ones.  That
transit was the first one available, just 24 h after its publication,
with an ingress and egress time  predicted by \citet{torres2010} at
2:16 and 4:28 UT of 2010 March 12.  A  $V = 9.39$ reference star lies
$4.4'$ from HAT-P-14.  The reference star has a color very similar
to the target.  We chose the full-frame readout in order to include a
third $V=10.53$ star to be used as a cross-check source.   They are
labeled \texttt{\#1} and \texttt{\#2} in the finding chart of
Fig. \ref{charts}, respectively.  The sky was photometric during the
entire series. Astronomical twilight was expected at 3:51 UT, but
observations lasted until the autoguide was possible, at nearly the
end of the nautical twilight. The series was stopped at 4:54 UT after
200 min without interruptions; 2247 frames  (in 4$\times$4-binning
mode)  were collected.  A high but quite stable seeing
($\sim$2.5$''$) and  a slight defocusing allowed for  constant 2 s
exposure  times  and a 37\% duty-cycle. In addition to the dome flats,
twilight flats were taken at dawn.

Finally, a complete transit of HAT-P-3b was observed  during the night
of 2010 April 7. HAT-P-3 is a $V = 11.86$ K dwarf, which hosts a
quite light (0.60 $M_{\rm J}$) transiting hot Jupiter with a short
period of $P\sim 2.9$ days \citep{torres2007}. The transit of HAT-P-3b
exhibits a depth that is twice as large (0.013 mag) as that of
HAT-P-14b.  The field is very poor, allowing only one star with a
significant flux ($V\sim 10.9$)  and a $B-R$ color not too different
from that of the target (1.4 versus 0.8), at about  $6.8'$. 
To reduce our overhead, we chose to read only a $90\times 500$
pixel CCD window  ($4\times 4$ binned). Seeing was excellent (FWHM
$\lesssim 1''$), so stars were  defocused at about $2''$ FWHM to
prevent saturation of the brighter one. A full time-series covering 70
min of pre-ingress and 60 min of post-egress was taken, with a
constant 5 s time sampling. The focus had be adjusted twice
during the series, because of the slow thermal expansion of the secondary
mirror.  In the end, 2882 frames were collected, with an overall
duty-cycle of 40\%. 

\section{Data reduction and analysis}
\label{reduc}

To maintain full control of the error budget and  
extract  the  maximal amount of information from the available data,
we implemented and used independent software tools   specifically
developed for this project.  The resulting pipeline, coded in Fortan
77/90 for the most part, is usable but still in active
development. As a crosscheck, its output is frequently compared with
that obtained using other publicly available softwares.

The key concept of our pipeline is the use of a fully empirical
approach to perform every task, from the  pre-processing, passing
through the light curve extraction, to the estimate of the transit
parameters along with their associated errors.  We care about keeping
our code as flexible as possible, so that  future  data sets coming
from a different instrumental setup  could  be treated with minimal
changes.

\subsection{Aperture photometry with \texttt{STARSKY}}
\label{starsky}

For our typical observing conditions (highly variable PSF due
to seeing and defocus, lack of stellar crowding), the method of choice
is differential aperture photometry,  which allows us to normalize the
flux of the target with a ``reference flux'' constructed by combining
one or more reference stars. Given the small size of our field, this
differential measure automatically cancels out first-order systematic
trends, such as transparency variations. We carry out this task with a
code named \texttt{STARSKY}. The  final aim of our empirical approach
is to obtain a light curve whose  \emph{measured} scatter is as small
as possible.

Each frame is calibrated using the supplied master-bias, -flat, and -dark.
These master frames are linearly interpolated when two given epochs
(before and after the observations) are available, accordingly to our
pre-reduction strategy.  A precise centroid and an estimate of
the FWHM of the stellar profiles is then computed by a Gaussian fit for
each target or reference star in an input list.  The sky level is
estimated over an annulus either as a  $k$-$\sigma$ clipped median or
by fitting a plane with an iterative $k$-$\sigma$ rejection.  The
$k$ factor is tunable, $k=3$ usually being the best choice.  The
annuli are dynamically adjusted for each frame $i$ by multiplying  the
input values of the inner and outer radii by the quantity $B_i/\langle
B_i\rangle$, that is the mean FWHM of the profiles, normalized by its
median over the full time-series.

After the sky is subtracted, fluxes are evaluated by summation over a
set of user-defined circular apertures; partial pixels are handled by
numerical integration of the fraction of pixel area falling inside
a given aperture. We considered ``optimal'' extraction (Naylor et
al. 1998) as a possible alternative to classical aperture photometry,
but the high spatial variability of the PSF  prevented us from such a
choice. Anyway, the advantages of that approach would have been partly
nullified by the very low contribution of the background noise to the
typical S/N of our sources.  \nocite{naylor1998}

\begin{figure}
\centering \includegraphics[width=9cm]{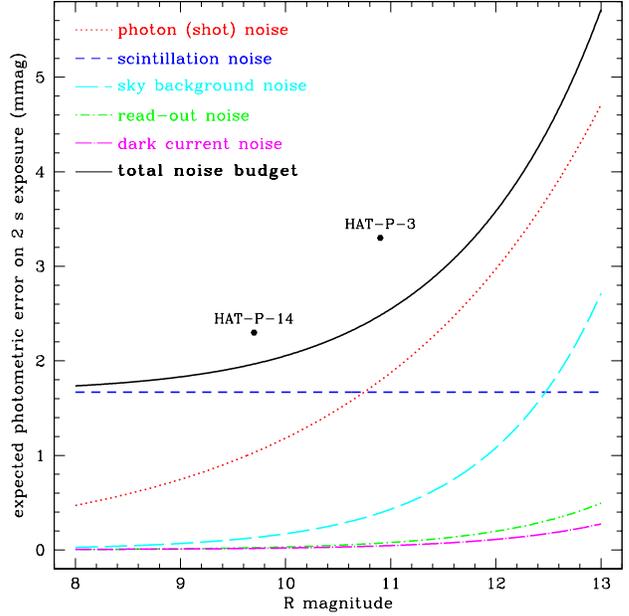}
\caption{Expected noise budget in mmag of a single 2 s exposure taken at the Asiago 1.82m 
telescope
under typical observing conditions (4$\times$4 binning, airmass 1.2, $7''$ aperture),
as function of the $R$ magnitude of the target. Measured off-transit scatters
are shown for the observed light curves of HAT-P-14 and HAT-P-3.}
\label{nb}
\end{figure}

The expected total  noise for each source is given
theoretically by \citet{merline1995} as
\begin{equation}\label{noise}
\sigma = \frac{1}{N_\star} \sqrt{N_\star + n_\textrm{pix}\left
  (1+\frac{n_\textrm{pix}}{n_{\rm B}}\right ) \left (N_{\rm S} + N_{\rm D} + N_{\rm R}^2
  \right ) + \sigma^2_{\rm s} } \quad\textrm{,}
\end{equation}
where $N_\star$ is the total number of photons collected from the
star, and $N_{\rm S}$, $N_{\rm D}$, and $N_{\rm R}$ the average  photons per
pixel coming  from the sky, the dark current, and the readout noise,
respectively. These quantities are assumed to have been estimated over
an aperture covering $n_\textrm{pix}$ pixels and  the sky over  an
annulus covering $n_{\rm B}$ pixels. The $\sigma_{\rm s}$ term is the
contribution from the atmospheric scintillation, given by
\citet{dravins1998} as
\begin{equation}
\sigma_s = 0.09 \, N_\star \frac{X^{3/2}}{D^{2/3}\sqrt{2t}}
\exp\left ( -\frac{h}{8}\right )\quad\textrm{,}
\end{equation}
where $X$ is the airmass, $D$ the diameter of the telescope in cm, $t$
the exposure time in s, $h$ the observatory altitude in km.  In the
typical observing conditions that we have  to deal with (bright targets,
short exposures, almost negligible dark current),  the noise budget is
largely dominated by photon ``shot'' noise and scintillation.
Figure~\ref{nb} shows how, in typical conditions, scintillation is
dominant for bright targets ($R\lesssim 11$), while shot noise is
dominant for faint targets ($R\gtrsim 11$). Read-out noise is nearly
negligible in our magnitude range.

Once we compute the ``absolute'' fluxes of each source along with the
associated expected errors, we need to combine the flux from the
reference stars to be used as optimal ``reference flux'' for our
differential photometry.   For each reference star $i$ of flux $F_i$
---and for each image---  we chose to calculate the reference flux
$F_{i}^0$  as the weighted mean of the  instrumental magnitudes of all
the other reference stars $F_{j\neq i}$.  This averaging technique 
was proven to be formally correct by \citet{broeg2005}.  During the first
iteration, the weights $w_i = 1/ \sigma^2_i$ are set according to  the
expected amount of noise given by Eq. (\ref{noise}).  A differential
light curve is constructed  by normalizing the raw flux $F_i$ to the
reference flux $F_{i}^0$.  The scatter $\tilde \sigma_i$ of each
differential light curve is estimated as  the $68.27^{\rm th}$
percentile of the residuals from its median value,  after the outliers
have been rejected with an iterative 4-$\sigma$ clipping
algorithm. Then we iterate the process of re-evaluating the reference flux $F_{i}^0$ 
by using this time the \emph{measured} scatters
$\tilde\sigma_i$ for the weights $w_i$,  constructing a new
differential light curve $F_{i}/F_{i}^0$, and so forth,  until
convergence is reached. 

To detect possible systematic trends or intrinsic variability
among the reference stars, two diagnostics are computed for each light
curve:
\begin{itemize} 
\item \texttt{dsys}, the ratio $\tilde\sigma_i/\sigma_i$ of the
  measured  photometric scatter to the expected amount of noise given
  by Eq. (\ref{noise}).\\
\item \texttt{psys}, the ratio of the $\tilde\sigma_i$ to the
  same scatter measured  after a low-order polynomial (whose degree is
  user-defined) has been  fitted and subtracted from the whole light
  curve. 
\end{itemize}
In the ideal case, i.e. without systematics and unknown sources of
random errors, we would get $\texttt{dsys}=\texttt{psys}=1$. A situation in which
$\texttt{dsys}\gg 1$, $\texttt{psys}\simeq 1$ is indicative of 
short-period systematics and/or an underestimated budget noise, while
$\texttt{psys}\gg 1$ reveals long-term  systematic trends (as these
might also depend, for example, on airmasses).  For the initial set of
reference stars, the reference flux $F^0_\star$ is computed by
combining all their fluxes and used to normalize the light curve
of the target $F_\star$, and the off-transit  scatter
$\tilde\sigma_\star$ is then evaluated.  The reference star(s) with the
highest values of \texttt{dsys} and \texttt{psys}  are removed from
the list of the valid reference stars and $F^0_\star$ is re-evaluated
without them, to check whether the off-transit measured
$\tilde\sigma_\star$ decreases. If so,  \texttt{dsys} and
\texttt{psys} are calculated for the remaining reference stars  and a
new selection is done, iterating until $\tilde\sigma_\star$ is
minimized.  Given the small number of reference stars in our fields,
each light curve is visually inspected to search for unexpected
behaviour.  

We incorporate an additional routine to perform empirical systematic
correction. We search for correlations  between the fluxes $F_i/F^0_i$ and relevant
quantities such as airmass, ($x$,$y$)  centroid positions, FWHM of the
stellar profiles and other relevant parameters, which we then 
fit with low-order polynomials and subtract. A systematic
correction is again considered effective if the out-of-transit
$\tilde\sigma_\star$ decreases when it is applied.  A further
cross-check is performed by sorting the stars by their $B-R$ colors to
see whether \texttt{dsys} and \texttt{psys} are higher for stars bluer or
redder than average, as we would expect in the case of color-dependent
systematics.

Finally the light curve of the target is normalized to one by dividing
it by its off-transit median flux, after a 4-$\sigma$ iterative
clipping, as explained above. The error bar associated with each
photometric point is assigned in this way: the expected noise for all
the stars  is computed using Eq.~(\ref{noise}), the error is then
propagated to the  reference flux for the target $F^0_\star$ and 
to the differential flux $F_\star/F^0_\star$.  This is an important
step, because often the error in the estimate of the reference
flux is not negligible when compared with that on the raw flux of
the target, especially when dealing with faint and/or an insufficient number of
reference stars.  Neglecting the uncertainties in the reference
sources, indeed,  ensures that $\sigma$ in Eq. (\ref{noise}) is
underestimated. The presence of red noise would also lead to an
underestimated $\sigma$.

\subsection{Light curve fitting} 

To extract the best estimate of the transit parameters,  we
need to fit a suitable model to our light curves. The details of the
fitting process can appreciably affect the results. It is even trickier 
to obtain a realistic estimate of the errors in the fitted
quantities, i.e. taking into account the role played by systematic
errors and input parameters that must be held fixed at a
constant value (above all, the limb darkening coefficients).  All the
light curves in this work were fitted with the JKTEBOP
code\footnote{\sf http://www.astro.keele.ac.uk/\textasciitilde{}jkt/codes/jktebop.html}
\citep{southworth2004}, which was originally developed for the analysis
of detached binary systems, of which planetary transits are a special
case; it has already been used in several works about high-precision
transit photometry \citep{southworth2008} and also specifically  for
TTV analysis \citep{coughlin2008}.  The package
models the two components of the system as biaxial ellipsoids and
performs a numerical integration in concentric annuli over the surface
of each component to estimate the flux coming from the system. 
It then fits the model to the data with a Levenberg-Marquardt (L--M)
least squares algorithm.  We chose JKTEBOP among other commonly used
modeling/fitting algorithms (such as those based on the analytic light
curves from Mandel \& Agol 2002) \nocite{mandel2002} because it does
not rely on small/spherical planet approximations, it converges
rapidly toward a reliable solution \citep{gimenez2006}, minimizes
the correlation between the fitted parameters \citep{southworth2008}
and includes several routines for the estimation of realistic
errors.

JKTEBOP can compute the limb darkening (LD) effect with several
different non-linear laws, as a simple linear dependence has been proven
to be not accurate enough  when dealing with high precision light
curves \citep{southworth2007}. We chose to use a quadratic LD law,
labeling with $u_1$ the linear term and $u_2$ the quadratic term. These two
quantities are strongly correlated and therefore difficult to fit
simultaneously; we instead usually fix $u_2$ to its theoretical
value, i.e. attempt to find only the $u_1$ term. In some cases this second
strategy may also deliver unphysical results for $u_1$ (e. g. negative
values) and so both $u_1$ and $u_2$ have to be fixed. We take the
theoretical LD coefficients by interpolating over the grid tabulated
by \citet{claret2000}  with the ATLAS model, using the $\log g$,
$T_{\rm eff}$, and [M/H] estimated for the host star from the highest
quality available spectroscopic follow-up. 

Since the internal errors in the L-M fit are known to be too
optimistic, JKTEBOP also includes some subroutines to estimate 
realistic uncertainties in every fitted parameter, using Monte
Carlo and bootstrapping methods \citep{southworth2008}.  We chose one
of the most conservative approaches, i.e. the  uncertainties given by a
``prayer bead'' empirical algorithm, also known as
``residual-permutation'' (RP).  The RP algorithm evaluates a best-fit
solution to the $N$ data points, and the residuals of the fit are then all
cyclically shifted to the next data point; a new fit is next found and the new
residuals are shifted again, this iteration continuing until the residuals return
to the original position (that is, after $N$ fits).  The errors
are then evaluated based on the final distribution of the fitted
parameters.  This algorithm automatically takes into account the
effects of both red noise and random errors (Jenkins et al. 2002). Its
main advantage over the other Monte Carlo Markov Chain and
bootstrapping methods is that it propagates over the light curve the
photometric features of the \emph{actual} correlated noise, without
being sensitive to the choice of an averaging time or to a specific
error-rescaling method. \nocite{jenkins2002} We also perform Monte Carlo
tests for comparison, to check whether  the errors estimated by the two
different methods are consistent with each other. In our analysis, we 
always choose the most conservative error. As shown in the discussion,
we performed thorough cross-checks to assess the reliability of the
parameters and the associated uncertainties derived by JKTEBOP.

We have already begun to implement from scratch an independent
pipeline that  will replace JKTEBOP for our purposes, i.e.\ following
the same flexible and empirical approach that characterizes
\texttt{STARSKY}. Once completed and tested, it will allow us to
perform the entire data reduction and analysis with fully customized
and optimized tools.

\subsection{HAT-P-14b} 

At $V=9.98$, $R\sim 9.7$, HAT-P-14 is the brightest star in the sample
surveyed by TASTE.  In our AFOSC field of view, only two stars provide
a non-negligible flux  compared to the target, and can be chosen as
primary reference stars: they are named \texttt{ref\#1}  ($R\sim9.1$)
and \texttt{ref\#2} ($R\sim 10.7$) in the right panel of
Fig. \ref{charts}.  In addition, light curves were extracted for another
eleven stars, with   $12.8\lesssim R \lesssim 16$ and
$0.6\lesssim(B-R)\lesssim 2.3$, just to check for  systematic errors
depending upon colors or upon ($x$,$y$) position.

For the inner and outer sky annuli we chose an initial value of 33
and 47 binned pixels, respectively (one 4$\times$4 binned pixel
$\simeq$ 1$''$), to be dynamically adapted to each frame as explained
in Sec. \ref{starsky}.  The aperture fluxes of the stars were
evaluated over 20 different circular apertures,   their radii being
equally spaced between 4.0 and 20.0 pixels.  The two available
master-bias  (taken just before and after the series) were
interpolated for each frame, while flat fielding was performed with a
constant master-flat taken at the morning twilight. We verified that
dark correction is not necessary -- in fact, its sole effect on
photometry is to add noise.  Measured background levels and raw
aperture fluxes confirm that the sky was photometric. The typical
gaussian FWHM of the profiles was quite stable through the series at
about 2.5$''$, because of a mix of atmospheric seeing and slight
defocus. The telescope guide drifted slowly and monotonically by about
1.6$''$ in 200 minutes, mostly along the RA axis.

\begin{figure}
\centering \includegraphics[width=9cm]{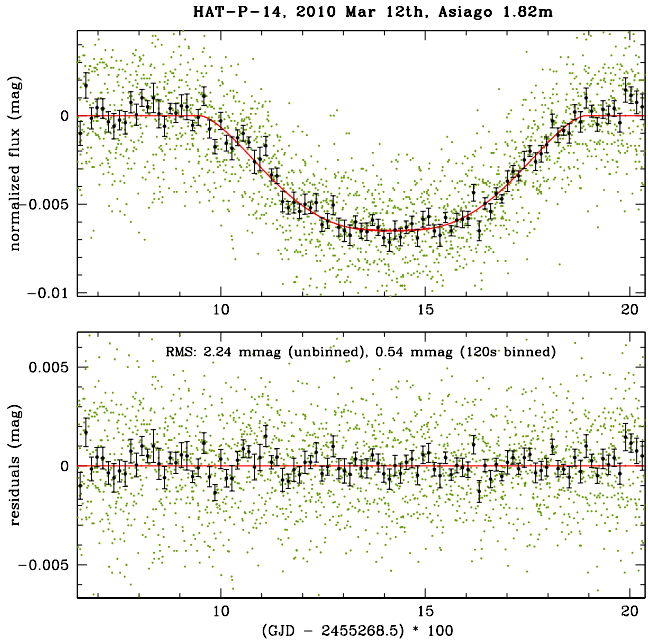}
\caption{\emph{Top:} 
Light curve for HAT-P-14b 
($V\sim 9.98$, $\Delta V = 0.007$), observed on March 13, 2010 
with the Asiago 1.82m. Unbinned points are shown in green and 120s-binned points in black.
Off-transit magnitude has been set to zero. \emph{Bottom:} residuals after the best-fit model 
is subtracted.}
\label{lchat14}
\end{figure}

The algorithms used for the iterative rejection of the reference stars
(described in Section \ref{starsky}) simplify in this case to only
three possible choices:  \texttt{1}, \texttt{1+2}, and \texttt{2}. When
using \texttt{ref\#2} as a single reference source, the off-transit
light curve of HAT-P-14b shows clear short- and long-term systematic
errors,  as confirmed by the \texttt{dsys} and \texttt{psys}
diagnostics. The brighter \texttt{ref\#1} has instead
$\texttt{dsys}\simeq\texttt{psys}\simeq 1$ for most of the aperture
sizes. We note that the color $(B-R)_{\tt \#1} = 0.7$
matches exactly the target, while $(B-R)_{\tt \#2} = 1.2$. To summarize,
\texttt{STARSKY} selected the flux  from \texttt{ref\#1} enclosed by a
7.2 pixel aperture as the best reference flux. The differential light
curve obtained for HAT-P-14 is shown with green points in the upper
panel of Fig. \ref{lchat14}.

Using JKTEBOP, we chose to fit the light curve by fixing both the LD
coefficients at their theoretical values of $u_1=0.33$ and $u_2=0.09$, which were
computed assuming the stellar parameters  given by \citet{torres2010} of
$T_{\rm eff} = 6600$ K, $v_{\rm mic}=0.85$ km/s,  $\log g = 4.25$, and
$\textrm{[Fe/H]}=+0.11$. Attempts to fit $u_1$ failed, as our single
light curve does not provide enough S/N for  such a shallow
transit. As soon as additional time series for HAT-P-14b are collected, a
simultaneous fit will allow us to get an empirical estimate for $u_1$
and $u_2$. 

The final fitted model is plotted as a red line in
Fig. \ref{lchat14}. When the model is subtracted from the data, the
unbinned light curve shows an RMS scatter of 2.2 mmag. Once the points
are binned to 120 s (i.e. $\sim 22$ frames per bin), the scatter
decreases to 0.54 mmag, showing only a small amount of red noise. The
errors do not show any significant correlation with airmass, seeing, or
positions, therefore no systematic correction was applied.  In spite
of an increase in the sky level of about two order of magnitudes, the
systematics were tightly constrained to within 1 mmag even in the last part
of the curve, when twilight occurred.  The fitted parameters are
reported in Table \ref{pars}, along with the  1-$\sigma$ errors
estimated by the residual-permutation algorithm.

\begin{table*}
\caption{Fitted parameters for HAT-P-14b (top panel) and HAT-P-3b (bottom panel).}
\label{pars}
\centering
\begin{tabular}{llll}
\multicolumn{4}{c}{\bf HAT-P-14b planet parameters} \\
\hline\hline
            & Torres10 &   & This work \\ \hline
$P$ (days)  & 4.627669    $\pm$ 0.000005  & &  4.627682 $\pm$ 0.000003 \\
$T_c$ (BJD) & 54,875.28938 $\pm$ 0.00047  & \phantom{54,856.70118 $\pm$ 0.00018}  & 55,268.64237 $\pm$ 0.00031  \\
$R_p/R_\star$ & $0.0805 \pm 0.0015$ & & $0.0834 \pm 0.0014$ \\
$b$ ($a\cos i/R$) & $0.891^{+0.007}_{-0.008}$ & & 0.93 $\pm$ 0.03 \\
$i$ (deg)   & 83.5 $\pm$ 0.3 & & $84.21 \pm 0.16$  \\
\end{tabular}

\vspace{0.5cm}

\begin{tabular}{llll}
\multicolumn{4}{c}{\bf HAT-P-3b planet parameters} \\
\hline\hline
            & Torres07 & Gibson10 & This work \\ \hline
$P$ (days)  & 2.899703    $\pm$ 0.000054\phantom{00}  & 2.899738     $\pm$ 0.000007 & 2.899737 $\pm$ 0.000004 \\
$T_c$ (BJD) & 54,218.7594 $\pm$ 0.0029    & 54,856.70118 $\pm$ 0.00018  & 55,294.56148 $\pm$ 0.00014  \\
$R_p/R_\star$& $0.1109^{+0.0025}_{-0.0022}$ & $0.1098^{+0.0010}_{-0.0012}$ & 0.1094 $\pm$ 0.0011 \\
$b$ ($a\cos i/R$) & $0.51^{+0.11}_{-0.13}$ & $0.576^{+0.022}_{-0.033}$ & 0.574 $\pm$ 0.018 \\
$i$ (deg)   & 87.24 $\pm$ 0.69 & $86.75^{+0.22}_{-0.21}$ & 86.75 $\pm$ 0.10 \\
\end{tabular}
\tablefoot{The columns show the orbital period $P$, the central instant of the transit
  $T_c$, the ratio of the radii $R_p/R_\star$, the impact parameter
  $b$, and the orbital inclination $i$.  Comparison with the previous
  estimates is given.  The reported 1-$\sigma$ errors are estimated by
  the residual-permutation algorithm. A constant amount of 2 400 000
  has been subtracted from all the BJDs.}
\end{table*}

\subsection{HAT-P-3b} 

As reported above, the field surrounding HAT-P-3 ($V=11.86$, $R\sim
10.9$, $B-R = 1.6$) is  extremely sparsely populated, containing only one suitable
reference star: TYC 3466-1158-1 ($R\sim 10.6$, $B-R = 1.4$).  The
narrow $1.5'\times 8'$ stripe we analyzed lacks in sources brighter
than $R=17$, leaving us with no other possible reference stars.  The inner
and outer sky annulus radii were set to 30 and 47 binned pixels,
dynamically adjusted throughout the series; we used 20 circular
apertures  for the star fluxes, equally spaced between 4.0 and 10.0
pixels. 

As for HAT-P-14, we interpolated the bias correction for each frame,
while a constant master-flat correction was applied, taken at the
morning twilight. The sky was photometric; the FWHM of the profiles
changed through the series in the range 1.5$''$-2.3$''$, because of the afore
mentioned thermal focus drift (Sec. \ref{observs}) and the
consequent adjustments. At the same time, the shape of the PSFs
changed noticeably, becoming asymmetric in the second part of the
series.  The telescope guide drifted monotonically by about 3.6$''$ in
250 minutes, along both axes. Reflecting the change of scale due to the
focus adjustments, we noticed a small differential drift, which caused
the relative distance between the two main stars to change by about
$0.8''$.

\begin{figure}
\centering \includegraphics[width=9cm]{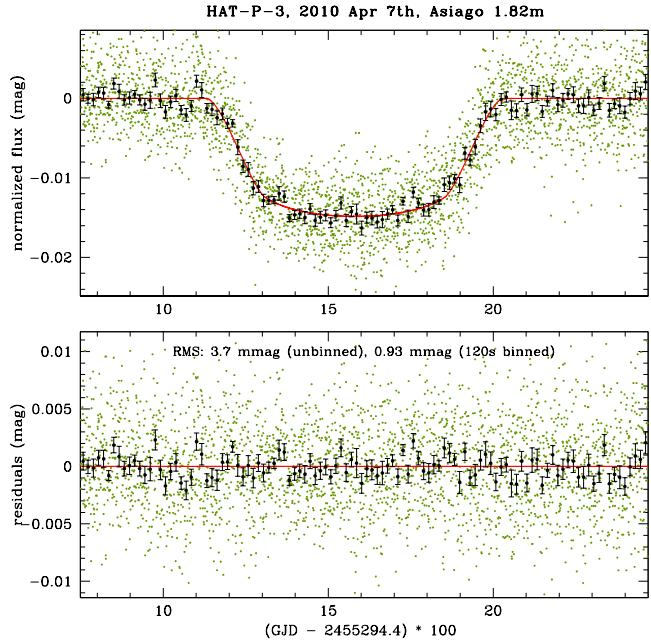}
\caption{\emph{Top:} 
Light curve for HAT-P-3b ($V = 11.86$, $\Delta V = 0.013$), observed 
on April 7, 2010 with the Asiago 1.82m telescope. 
Unbinned points are shown in green and 120s-binned points in black.
Off-transit magnitude has been set to zero. \emph{Bottom:} residuals after the best-fit model 
is subtracted.}
\label{lchat3}
\end{figure}

TYC 3466-1158-1 turned out to be a good choice, showing no long-term
systematic trends ($\texttt{psys}\simeq 1$) similar to those expected from
differential extinction. However, $\texttt{dsys}=1.24$ indicates that there is a
non-negligible level of short-term systematics, probably owing to small
pixel-to-pixel residual errors in the flat-field correction, combined
with the PSF unpredictable (also spatial-) variability, which led to
changes in the photometric zero point.  An intrinsic microvariability
of TYC 3466-1158-1 is a possible cause that should not be ruled
out.  In addition, a part of the S/N that would be achievable on
HAT-P-3 is inevitably lost because of the paucity of reference flux.
\texttt{STARSKY} selected 6.1 pixels as the best aperture. The
differential light curve obtained for HAT-P-3 is shown with green
points in the upper panel of Fig. \ref{lchat3}.

The light curve was fitted by fixing both the LD coefficients at their
theoretical values $u_1=0.47$, $u_2=0.24$, which werecomputed assuming the
stellar parameters given by \citet{torres2007} of $T_{\rm eff} = 5185$
K, $v_{\rm mic}=2.0$ km/s,  $\log g = 4.61$, and
$\textrm{[Fe/H]}=+0.27$. Attempts to fit $u_1$ failed, as our single
light curve does not provide enough S/N.  The final fitted model is
plotted as a red line in Fig. \ref{lchat3}. When the model is
subtracted from the data, the unbinned light curve shows an RMS
scatter of 3.7 mmag. Once the points are binned to 120 s (i.e. $\sim
24$ frames per bin), the scatter decreases to 0.93 mmag. Though a
small amount of systematic noise was detected, it does not show a
significant correlation with airmass, seeing, or positions, therefore
no correction was applied.  As for HAT-P-14, the increase in the sky
level near the end of the series does not cause significant trends:
even in the last few minutes, the binned points are well within 2-$\sigma$
off the model.  The fitted parameters are reported in the Table
\ref{pars}, along with the 1-$\sigma$ errors estimated by the
residual-permutation algorithm.

\subsection{TTV analysis}
\label{ttvan}

When TASTE begins to accumulate multiple transits for a given target, a
search for TTV will be carried out using specially developed software
tools, now still at the development stage.

In principle, at least three measurements are needed to detect a
TTV. In a ``real world case'', the signal could be multimodal, noisy,
and/or unevenly sampled, and the number of observed points required
for a TTV detection is strongly dependent on the signal itself.
Previous works (Gibson et al. 2009, among others)  have shown that it
is risky to rely only on unexpectedly large amounts of scatter in the
($O-C$) diagrams.  The solid detection of a TTV requires instead the
identification of periodic structure in the timing residuals. For
typical targets, a number of transits between 5 and 10 is sufficient
to place very stringent upper limits on the system. The TTV signals  of
Kepler-9b ($\sim 2$ min) and Kepler-9c ($\sim 4$ min) are the only
ones detected so far (Holman et al. 2010), which were sampled,
respectively, with 6 and 9 observed transits with a timing accuracy of
about 80 s (that is, working at S/N$\sim 2$--3).

If the $O-C$ residuals for a given TASTE target show an anomalous
scatter (reduced $\chi ^2 \gg 1$), possible periodicities will be
searched for by a periodogram, folding the residuals over the significant
frequency peaks. Even in the case of null detection, upper limits will be
estimated by fitting synthetic TTV signals to our data. These will be
calculated by integrating the equations of motion on a grid of
eccentricities, masses, and periods of the hypotetical perturber and
mapping the $\chi ^2$ over the parameter space. 

\section{Discussion}

The overall photometric performances of our system met our
expectations. The scatter measured in the whole light curve of our
targets is in good agreement with the theoretical expected amount of
noise, as shown in Fig. \ref{nb}.  This holds both for  HAT-P-14 and
HAT-P-3, which are, respectively,  the brightest, and close to the
faintest magnitude limits of our surveyed sample.  In particular, the
noise measured for HAT-P-14 ($\sim 0.5$ mmag on 120 s bins) is indicative of
a very small amount of systematic error, and of the same order
as achieved by state-of-the-art photometry with medium class telescopes
(e.g., see Southworth et al. 2009b, with data acquired in a much
better site).  

\begin{figure}
\centering \includegraphics[width=9cm]{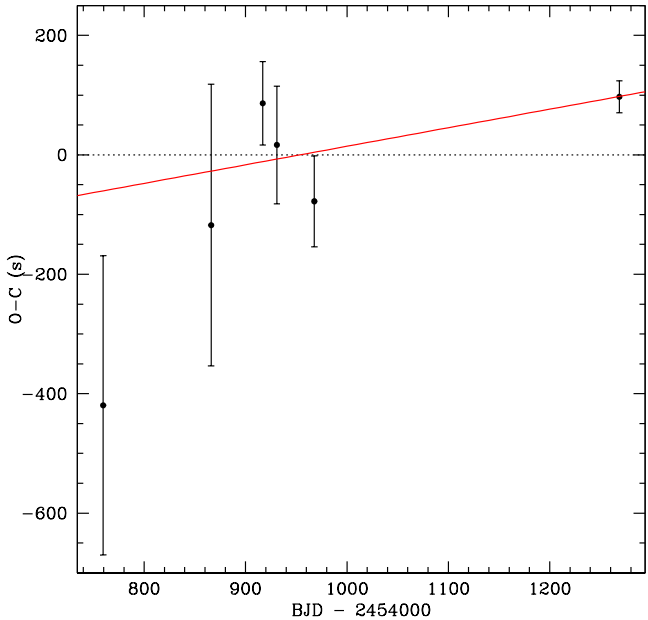}
\caption{$O-C$ diagram for the central time of the 
HAT-P-14b transit. The last point is from TASTE (2010, Mar 12),
the others from \citet{torres2010} from which the $C$ ephemeris is taken. 
The red line is the weighted fit for our refined ephemeris.}
\label{ttv14}
\end{figure}

HAT-P-14b was very recently discovered and no specific TTV studies for it have yet
been performed. The shallowness and grazing shape of
its transit make it one of the most difficult targets in our sample.
The $O-C$ diagram (observed minus computed) for the transit central
times, relative to the ephemeris published by \citet{torres2010} is
shown in Fig. \ref{ttv14};  we plotted the five measurements made by
\citet{torres2010}, as well as the first TASTE point. The improvement
with respect to the original data and due to our 25 s timing accuracy
is remarkable.  Because of the better accuracy and the longer than one-year
baseline, we were able to improve  the ephemeris  by using
all the weighted available points.

Our best-fit value for the relative radius $R_p/R_\star$ of HAT-P-14b
is $\sim 4\%$ larger than that reported by \citet{torres2010},  though
the two estimates are consistent to within 1-$\sigma$.  Despite the
lower scatter in our light curve, we cannot so far improve the
precision in $R_p/R_\star$ over the previous measure, which is based
on a simultaneous fit of five transits.  We instead refined the other
orbital parameters, which are in good agreement with
\citet{torres2010} except for our derived inclination $i$, which is
slightly larger \footnote{After the submission of the present paper, an
  additional study of HAT-P-14b appeared on \texttt{astro-ph}
  \citep{simpson2010}.}.

\begin{figure}
\centering \includegraphics[width=9cm]{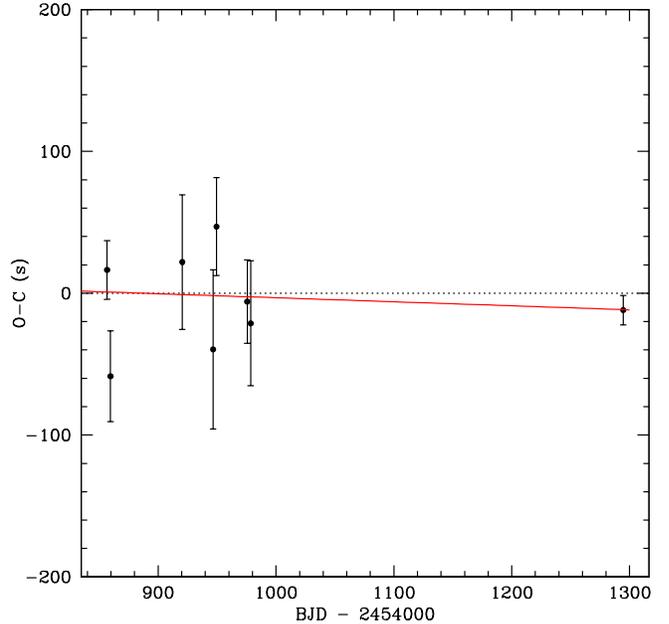}
\caption{$O-C$ diagram for the central time of the HAT-P-3b
transit. The last point is from TASTE (2010, Apr 7), the others from
\citet{gibson2010} from which the $C$ ephemeris is taken.  The point
by \citet{torres2007} is omitted.  The red line is the weighted fit
for our refined ephemeris.}
\label{ttv3}
\end{figure}

As for HAT-P-3b, a TTV null detection was found recently, and published by
\citet{gibson2010}, placing a constraint on planets  with masses higher
than 0.33 and 1.81 $M_\oplus$ (in inner and outer 2:1 resonances,
respectively). The very accurate ephemeris that they calculated (which
accumulated only $\sim 90$ s of uncertainty at our epoch) allows us to
cross-check very strictly the timing  accuracy we reached. Our
estimated error ($\sim 11$ s) is smaller by a factor 2--5 than the
RISE single measurements. Our measure is indeed well  within the
range predicted by the previous ephemeris, as can be seen from
Fig. \ref{ttv3}.  Our improved ephemeris  is therefore nearly
unchanged except for the smaller error on both $P$ and $T_c$.  The
other parameters are in excellent agreement with the results by
\citet{gibson2010}, with uncertainties that are similar overall.

\begin{figure}
\centering \includegraphics[width=9cm]{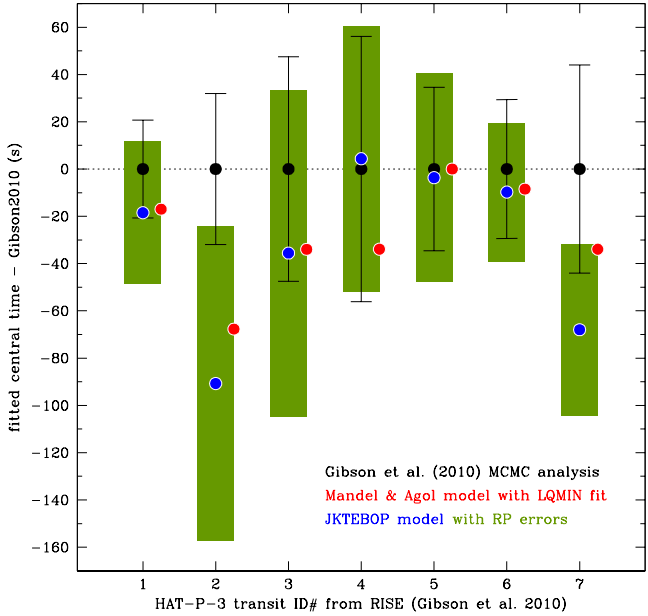}
\caption{Central times measured on seven light curves of
    HAT-P-3 collected by Gibson et al. (2010).  Comparison between the
    values published in the original study (black circles and error
    bars) and those found by our re-analysis with JKTEBOP (blue
    circles with green error bars from  RP algorithm) and a
    Mandel \& Agol model (red circles, slightly displaced on the right
    for clarity). The numbering of the RISE transits is consistent
    with Fig. \ref{gibson}.}
\label{crosscheck}
\end{figure}

\begin{figure}
\centering \includegraphics[width=8.5cm,trim=40 30 30 30]{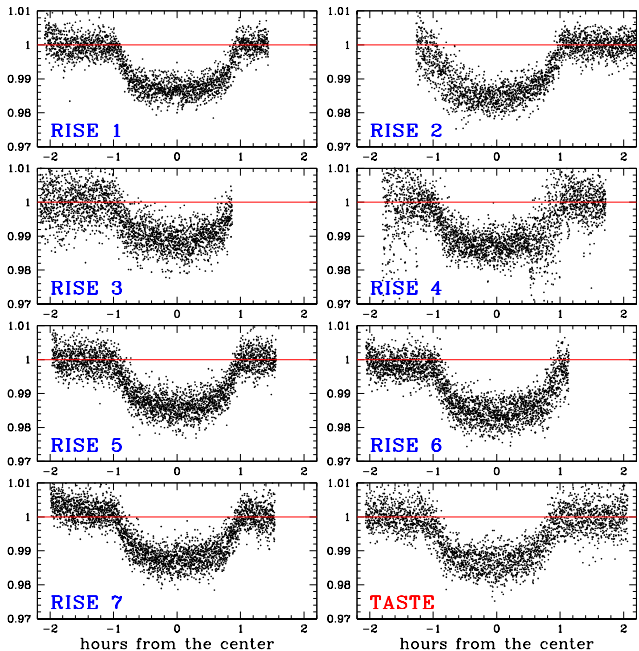}
\caption{Raw light curves for seven transit of HAT-P-3b by Gibson et al. (2010) (RISE\#1, \dots, RISE\#7)
and the TASTE transit shown in Fig. \ref{lchat3}. The light curves are shown in the same scale and without
binning.}
\label{gibson}
\end{figure}

To assess the reliability of our analysis, we performed a
cross-check by analyzing with JKTEBOP the seven light curves of
HAT-P-3b used by Gibson et al. (2010) in their TTV study. We corrected
the raw light curves by fitting the out-of-transit flux with a linear
function, as described by the authors. We then fixed both the LD
quadratic coefficients at the values used in that study and ran
JKTEBOP by estimating the errors with the RP algorithm.  As an
additional test, we also determined the central times by fitting a
Mandel \& Agol model with a L--M least squares algorithm. The results
are shown in Fig. \ref{crosscheck}, where  we plotted the difference
between our timings and those published, for each of the seven RISE
transits. Our measurements are fully consistent with the results of Gibson
et al. (2010), as are the associated error bars. The three
transits that display the largest deviation are problematic in various ways
(Fig. \ref{gibson}): two are partial transits with a strong variation
in  scatter during the series (RISE \#2, RISE \#3), while the  third
displays clear systematic errors in the out-of-transit part (RISE \#
7). Even the highest quality RISE curves (\#1, \#5), despite their
smaller scatter, display an amount of red noise that is larger than that
detected on the TASTE transit, as can be seen by binning each curve
and comparing the resulting RMS with the one expected by a
$\sqrt{t_\textrm{exp}}$ scaling law.

\section{Conclusions}

The primary aim of the TASTE project is to collect a library of
high-precision, short-cadence light curves for a selected sample of
transits, by performing differential photometry at the Asiago 1.82m
telescope. The focus of the project is on the search for TTV and TDV
effects induced by other planets (or moons) in the limited number of
monitored systems. An important additional advantage is the opportunity to
perform a long-term photometric follow-up, leading to an extremely
precise characterization of each target, as done e.g. by the TLC
project \citep{holman2006}. By keeping in mind both of the
above-mentioned goals, a careful selection of the targets was
performed, as described in Section \ref{sample}.

Many improvements are still possible, above all increases in the
duty-cycle, which was only $\sim 40\%$ for the two time series that have been
presented in this work.  Recent tests have demonstrated that, when reading a
narrow CCD window, a further reduction  of the technical overhead from
3--3.5 s to about 1.6 s is possible. This would lead to a $\gtrsim
70\%$ duty-cycle (for 4 s exposures) for most of our surveyed targets,
decreasing the photometric scatter on average by 30--35\%.  The
real-time computing of our photometry will also allow us to perform a
nearly-instant correction of the small tracking drifts that we mentioned in
Section \ref{reduc}, minimizing any residual systematics caused by an
imperfect flat-field correction. 

The fine tuning of our instrumental setup will proceed in step with
the additional improvement of our photometric code and the
implementation of a brand new independent software tool for the
light curve analysis.
However, the first results reported here already show the feasibility
of our project.  We have already began to collect light curves for two
targets in our sample (HAT-P-14b and HAT-P-3b), which will define the
``zero epoch'' of our analysis.  In both cases, the achieved
photometric precision is very close to the theoretical limits, and the
systematics are constrained to well within the milli-magnitude.  The
overall measured scatter is close to that of the present optimal value achievable
1-2 m class telescopes, as obtained by long-cadence  photometry on
similar targets. The timing accuracy that we have reached, estimated by JKTEBOP
algorithms, demonstrates that TASTE is very competitive compared to the
performances of other similar projects. With only one  transit
available  for each target, we have been able to derive a refined
ephemeris and an independent parameter estimate for both targets.

\begin{acknowledgements}
We thank M. Fiaschi for the support in the optimization for our
purposes of AFOSC data acquisition software, and L. Malavolta for the night support
during the observations of HAT-P-14b and HAT-P-3b.

\end{acknowledgements}

\bibliographystyle{aa}
\bibliography{15199}

\end{document}